\documentclass[reprint, amsmath,amssymb, aps ]{revtex4-1}

\usepackage[pdftex]{graphicx}
\usepackage{amsmath}
\usepackage[caption=false,font=normalsize,labelfont=sf,textfont=sf]{subfig}
\usepackage{url}
\usepackage[mode=text]{siunitx}
\usepackage[colorlinks=true, linkcolor=blue, citecolor=blue, urlcolor=blue, bookmarks=true, breaklinks=true]{hyperref}
\usepackage{bookmark}
\usepackage{tikz}
\usetikzlibrary{positioning}
\usetikzlibrary{patterns}
\usetikzlibrary{spy}
\usepackage{pgfplots}
\usepgfplotslibrary{units}
\pgfplotsset{width=10cm,compat=1.9}
\usepackage{nicefrac}

\labelformat{equation}{eq.~(#1)}
\def\equationautorefname#1{}
\begin{document}
\title{Simulation of short pulse photoemission in \\ a micro-diode with implications for \\ optimal beam brightness}%
\author{H\'akon~\"Orn~\'Arnason,
        Kristinn~Torfason,
        Andrei~Manolescu,
        and~\'Ag\'ust~Valfells}%
\affiliation{Department of Engineering, Reykjavik University, Menntavegur 1, IS-102 Reykjavik, Iceland.}



\begin{abstract}
Molecular dynamics simulations, with full Coulomb interaction are used to model short-pulse photoemission from a finite area in a microdiode. We demonstrate three emission regimes, source-limited emission, space-charge limited emission for short-pulses, and space-charge limited emission for the steady state. We show that beam brightness is at a maximum during transition from the source-limited emission regime to the space-charge limited emission regime for short pulses. From our simulations it is apparent that the important factor is the emitter spot size when estimating the critical charge density for short-pulse electron emission.
\end{abstract}

\maketitle

\section{Introduction}

Short pulse electron beams are important in many applications, e.g.\ high power microwave sources for hundreds of 
GHz and THz ~\cite{booske_plasma_2008}, time resolved electron microscopy~\cite{zewail_4d_2010,sun_direct_2020}, and free electron lasers~\cite{barletta_free_2010}.
Ideally these bunches should be coherent and have high current, a characteristic which can be quantitatively measured in terms of the beam brightness.
In this paper we use the following definition of brightness from Reiser~\cite{reiser_theory_2008}, 

\begin{equation}
    Br = \frac{\eta I}{\epsilon_x \epsilon_y} \propto \frac{I}{\epsilon^2}\ ,
    \label{eq:Brightness}
\end{equation}
where $\eta$ is a geometric constant equal to $2/\pi^2$, $I$ is the current along the beam direction, $z$, and $\epsilon_{x,y}$ are the emittances in the transverse directions, $x$ and $y$, describing the lateral spread of the beam in the phase space~\cite{Edwards_book,Jensen_book}, which are expected to be equal for a beam with transverse symmetry. 

High current and low emittance are somewhat competing goals, as high current beams are subject to space-charge forces that lead to increased emittance and also because high current generally implies high electron density near the cathode that leads to increased scattering and consequentially higher emittance~\cite{zhang_spacecharge_2021}.
This trade-off between current and emittance suggests that an optimal value of brightness exists.
In fact, this has been observed for pulsed photoemission, both experimentally~\cite{bazarov_maximum_2009}, and from simulations~\cite{kuwahara_boersch_2016}.
Optimal brightness is not limited to pulsed photoemitted beams.
For instance, it has been observed in simulations of thermal emission in microdiodes, that optimal brightness is obtained during the transition from source limited emission to space-charge limited emission~\cite{sitek_space-charge_2021}.
The work presented in this paper was initiated as a study of the corresponding transition to space-charge limited photoemission from a planar cathode. 

For that purpose we use a high-fidelity molecular dynamics code we have developed, called Reykjavik University Molecular Dynamics for Electron Emission and Dynamics (RUMDEED)~\cite{sitek_space-charge_2021,torfason_molecular_2016}, including discrete particle emission, scattering, and space-charge effects on electron emission and propagation, to model the physics of electron beamlets near the cathode.
This includes emission, propagation and effects on derived parameters such as pulse charge, emittance, and particularly brightness.
We investigate the transition from source limited to space-charge limited emission and show how optimal brightness can be obtained with regard to nonlinear physics in the vicinity of the cathode. 

There are a number of sources of emittance growth that can diminish the brightness of a beam as it propagates through a device due to effects such as misalignment of focusing and accelerating components, beam mismatch, nonlinear forces, etc.~\cite{reiser_theory_2008}.
However, it is useful to look carefully at what is happening at the cathode and in its immediate neighborhood for a better understanding of ultimate limits to brightness.
The purpose of this paper is to isolate space-charge and discrete particle effects on photoemission of electrons and their propagation in the immediate vicinity of the cathode, how the transition from source limited emission to space-charge limited emission happens, and to show how that affects the beam brightness.
To do so we use a model that does not incorporate cathode temperature, surface protuberances, or variable work function on the cathode surface.
We will also compare our results on the transition from source limited emission to space-charge limited emission to commonly used models.

\section{Methodology}

In this paper the molecular dynamics computer simulation (MDCS) method is being used to investigate how the electron beam evolves with regards to emitter size, width and amplitude of the photon pulse, and applied potential.
The MDCS method is well suited for a system with a relatively small number of particles.
Our system is an infinitely wide vacuum diode, with the anode-cathode gap spacing denoted by $D$, zero voltage at the cathode, and applied voltage $V$ at the anode.
The emission area is a disk with radius $R$ which is smaller than $D$.
The work function of the emitter $\phi$ is uniform over the area and equal to average energy of the photons. 

The photons in a pulse have energies $E$, with a Gaussian distribution, with average $\langle E \rangle = \hbar\omega$, and with a very small standard deviation $\sigma_\mathrm{E}\ll \hbar\omega$.
The number of photons $N$ within the pulse have also a Gaussian distribution, but as a function of time.
The time is discretized in small steps $\delta t$, the width of the pulse is $\sigma_\mathrm{N}$ time steps, and in practice the total duration of the pulse $\tau_\mathrm{p}$ is assumed equal to $16\sigma_\mathrm{N} \delta t$.  

The electrons are emitted with the initial velocity $v_{z0}$, corresponding to the excess energy transferred by the incoming photon and the emission occurs in the direction normal to the surface of the emitter,
\begin{equation}
    v_{z0} = \sqrt{2(E - \phi) / m } \ ,
    \label{eq:initial_velocity}
\end{equation}
where $m$ is the electron mass.
The dispersion of $v_{z0}$, due to the small dispersion of the photon energy, is obviously also small.
The number of emitted electrons is a function of time, and it follows the Gaussian photon pulse.
At each time step the theoretical (or virtual) number of electrons that could possibly be emitted is modeled as a Poisson random variable, with an expected value given by the quantum efficiency (QE) multiplied by the number of incoming photons during that time step, which in fact is the average number of emitted electrons in the absence of an electric field and space-charge effects~\cite{jensen_photoemission_2009}.
Although the pulse duration is short, below 
1 ps, we neglect possible multiphoton emission events.
In our numerical implementation the pulse intensity is described by the amplitude of the virtual electron pulse emitted by the cathode (i.e.\ the mean value at the middle of the pulse), which is an input parameter that we refer to as the scaled laser amplitude.  
\newcommand{\yslant}{0.4}
\newcommand{\xslant}{-0.85}
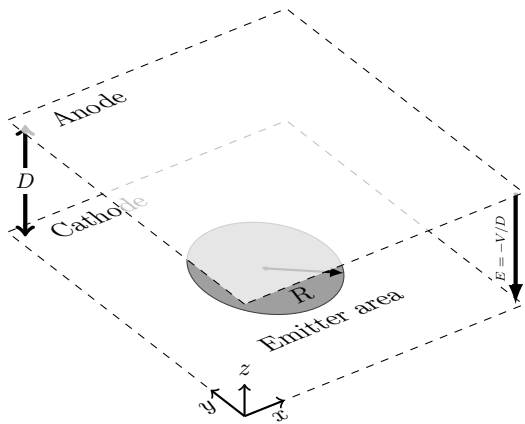
\begin{figure}
    \centering
    \begin{tikzpicture}[scale=0.53]
        \tikzstyle{ann} = [fill=white,font=\footnotesize,inner sep=2pt]
      \begin{scope}[
          yshift=-80,
          every node/.append style={yslant=\yslant,xslant=\xslant},
          yslant=\yslant,xslant=\xslant
        ]
        \draw[black, dashed, thin] (0,0) rectangle (7,7);
        \fill[black]
        (0.5,6.5) node[right, scale=1] {Cathode};
        \draw[fill=gray, opacity=.75]
        (3.5,3.5) circle (1.5); 
        \draw[fill=black, opacity=.75]
        (3.5,3.5) circle (.05); 
        \draw[-latex, thick](3.5,3.5) to (4.7,2.6);
        \draw (3.5,2.9) node[below] {$R$};
        \draw (3.5,2) node[below] {Emitter area};
        \draw[thick,->] (0,0,0) -- (1,0,0) node[anchor=north east]{$x$};
        \draw[thick,->] (0,0,0) -- (0,1,0) node[anchor=north east]{$y$};
    
      \end{scope}
    
      \draw[-latex, ultra thick](6.8,2.8) to (6.8,0);
      \draw (6.7,1.4) node[ann, above, rotate=90,scale=0.6] {$E = -V/D$};
      \draw[arrows=<->, ultra thick](-5.5,1.7) to (-5.5,4.5);
      \node[ann] at (-5.5,3.1) {$D$};
      \draw[thick,->] (0,-2.8,0) -- (0,-2,0) node[anchor=south]{$z$};
    
      \begin{scope}[
          yshift=0,
          every node/.append style={yslant=\yslant,xslant=\xslant},
          yslant=\yslant,xslant=\xslant
        ]
        \fill[white,fill opacity=.75] (0,0) rectangle (7,7);
        \draw[black, dashed, thin] (0,0) rectangle (7,7);
    
        \fill[black]
        (0.5,6.5) node[right, scale=1] {Anode};
      \end{scope}
    \end{tikzpicture}
    \caption[]{Simulated vacuum diode system model, circular emitter area with radius $R$, gap size $D$ is 
2500 nm and applied potential $V$.}
    \label{fig:system}
\end{figure}
The mechanism of electron ejection from the cathode including the space-charge effect due to the already emitted electrons is implemented in our code and was used in previous work~\cite{torfason_molecular_2015, torfason_high-fidelity_2018, torfason_molecular_2016, haraldsson_molecular_2020, gunnarsson_space-charge_2021,Torfason_dynamics_2021}.
The algorithm selects an emission site randomly within the circular emitter area of the cathode.
First, the energy of the photon is compared to the work function at the site to see if emission is possible.
Then, if the electric field at that site favors the emission, i.e.\ it is oriented such that the force on the electron would accelerate it towards the anode, a second check is made at 
1 nm above the cathode, and if the field there is also favorable, an electron is placed 
1 nm above the cathode surface and given velocity according to \autoref{eq:initial_velocity}.
If neither the work function or the local electric field at the candidate site is favorable, no electron is placed outside the cathode, and a failure of the emission event is recorded.
This process is repeated until either the maximum number of emitted electrons allowed by the Gaussian pulse at that time step, or \num{100} recorded failures to place, have been reached.
With the Gaussian limit not being reached, or ultimately with no favorable site for emission, will indicate the space-charge limited regime.
This mode of electron placement ensures a self-consistent current density over the emitter cathode, whether source limited or space-charge limited.

The next step of the simulation is to calculate the net forces acting on every electron due to the direct Coulomb interaction with the other electrons in the system, including the image charge partners outside the boundaries of the simulation box, and due to the accelerating field created by the anode.
Then the time step is advanced and the Velocity-Verlet algorithm is used to calculate the new positions and velocities of the electrons.
Electrons that pass the boundaries of the anode or cathode are recorded and removed from the system.
The simulation works through the process of electron emission and advancement for each time step until a user selected end point.
The total current through the diode is calculated using the Ramo-Shockley theorem~\cite{ramo_currents_1939,shockley_currents_1938},
\begin{equation}
    I = \frac{q}{D}\sum_i v_{iz} \ ,
    \label{eq:Ramo_current}
\end{equation}
where $q$ is the electron charge, and $v_{iz}$ the component of the instantaneous velocity of the electron $i$, that is normal to the cathode surface (i.e.\ in the $z$ direction).

In order to observe the effect of pulse relatively to the entire system size, we normalize the pulse width $\tau_\mathrm{p}$ with the transit time of a single electron from the cathode to the anode, 
\begin{equation}
    \tau = D \sqrt{\frac{2 m}{q V}} \ ,
\end{equation}
yielding $\tau_\mathrm{n}=\tau_\mathrm{p}/\tau$, or
\begin{equation}
    \tau_\mathrm{n}  = \frac{16\sigma_\mathrm{N}\delta t}{D} \sqrt{\frac{V q}{2 m}} \ .
    \label{eq:pulse-norm}
\end{equation}
Thus, if the duration of the laser pulse is equal to the time it takes for a single electron to be accelerated from the cathode to the anode then $\tau_\mathrm{n}=1$.

In the interest of calculating the brightness defined in \autoref{eq:Brightness}, we use the maximum value of the Ramo-Shockley current given by \autoref{eq:Ramo_current}, and the statistical emittance, appropriate for our computational approach~\cite{jensen_general_2007},
\begin{equation}
    \epsilon_x = \sqrt{ \langle x^2 \rangle \langle x'^2 \rangle - \langle xx' \rangle ^2} 
    \label{eq:emittance_x}
\end{equation}
with $x$ denoting the position and $x'=dx/dz=v_x/v_z$ being the deviation angle of the particle in the $x$ direction.
The angular brackets represent averages over all electrons when they pass through the anode.
The corresponding similar formula is used for the $y$ direction.

The numerical values of the parameters used in the simulations are: 
The radius of the disk-shaped emission area of 
125, 187.5, 250 nm; 
the distance between the cathode and the anode $D=2500$ nm; the anode-cathode potential difference of 50, 75, 100 Volt; the simulation time step $\delta t=0.25$ fs with the total running time of 15 ps; the center of the emission pulse 5 ps with amplitude 
1, 2.5, 5, 10, the average number of electrons at emission peak and pulse width from 
4 fs tp 4000 fs; mean energy of photons 4.7 eV with a standard deviation of 
0.02 eV; the work function of the material was chosen to be 4.7 eV, approximating copper as cathode material.

\section{Results and analysis}

\begin{figure}[!t]
    \centering
    \includegraphics[width=\linewidth]{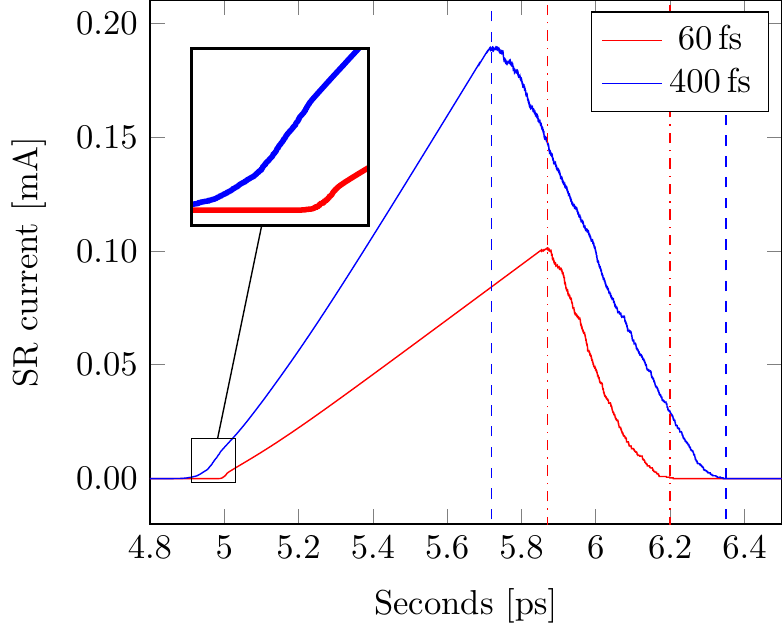}
        \caption{Temporal profile of induced current for two different laser pulse widths (400 fs in blue, and 60 fs in red). Gap voltage is 75 V, emitter radius of 250 nm, Scaled laser amplitude of 10. Dashed and dashdot lines indicate when first and last electrons cross anode.}
        \label{fig:Ramo_current}
\end{figure}

We begin by looking at the current induced in the diode as a function of time, as shown in \autoref{fig:Ramo_current}.
Here we see different values of the pulse width, $\tau_\mathrm{p}$, while the gap voltage is fixed at 75 V, and the emitter radius and amplitude of the laser pulse are held constant as well.
Recall that the peak of the laser pulse is located at time $t =5 $ ps, and note that the transit time for a single electron across the diode gap for the given voltage is 
$\tau= 0.97$ ps.
In the case of $\tau_\mathrm{p}= 60$ fs, emission is source limited and the charge is emitted in a tight bunch at approximately 5 ps.
This bunch of charge is subsequently accelerated across the gap by a nearly constant applied electric field, resulting in an induced current that grows linearly with time until the foremost electrons in the bunch are absorbed by the anode, at which time the induced current begins to drop due to absorption of charge.
In the case of $\tau_\mathrm{p} =400$ fs, emission is space-charge limited and begins slightly prior to the 5 ps mark but nonetheless manifests as a bunch of electrons that are accelerated across the diode gap leading to linear growth of the induced current in time until the leading electrons are absorbed at the anode.
From the figure it is apparent that the bunch length is greater for the wider laser pulse as would be expected.

\begin{figure}[!t]
    \centering
    \includegraphics[width=\linewidth]{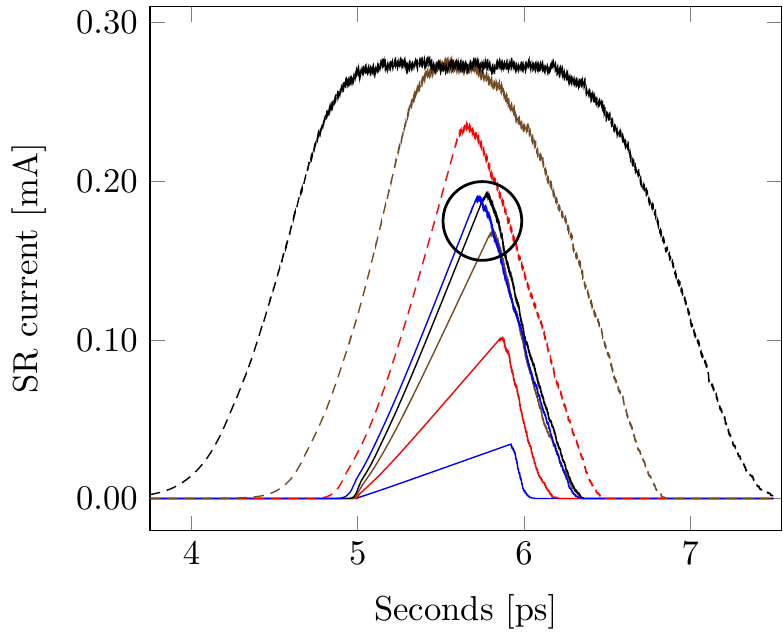}
        \caption{Temporal profile of induced current for a number of different laser pulse widths ranging from 20 fs to 4000 fs. Gap voltage is 75 V, emitter radius of 250 nm, Scaled laser amplitude of 10. The lowest peak current is on the trace for smallest pulse width. The circle surrounds peaks of traces for pulse widths ranging from 120 fs to 400 fs.}
        \label{fig:multi_ramo_current}
\end{figure}

In the conventional analysis of the Child-Langmuir current it is implicit that the current is constant in time for fixed external parameters~\cite{zhang_100_2017}.
As the width of the laser pulse is increased so as to exceed the transit time across the gap, we see a transition in the current profile as a function of time when looking at \autoref{fig:multi_ramo_current}.
For long values of the pulse length the rise in the current is no longer linear, as one would expect for a short electron bunch, but has a steadily increasing gradient due to the fact that the number of electrons being emitted from the cathode increases gradually with time.
We also observe a plateau in the current which occurs once the diode gap has been filled with a steady current. This plateau corresponds to the steady-state space-charge limited current in the diode.
Before looking in more detail at how the normalized pulse length affects the current it is worthwhile to look at some common models for analysis of space-charge limited emission from finite area emitters and for short pulses.
Lau~\cite{lau_simple_2001} devised a simple and elegant theory for steady-state space-charge limited current from a finite emitting area which was later extended by Koh and Ang~\cite{koh_3DCL_2005}.
From Koh and Ang's work it is expected that the current density from a circular emitter of radius, $R$, and diode gap spacing, $D$, when the emission energy is negligible should be:
\begin{equation}
    J_\mathrm{2D} =J_\mathrm{CL}\left(1 + \frac{D}{4R}\right)\, ,
    \label{eq:Koh}
\end{equation}
where $J_\mathrm{CL}$ is the classical Child-Langmuir current for an infinite planar diode, given by 
\begin{equation}
    J_\mathrm{CL} =\frac{4\pi}{9}\varepsilon_0\sqrt{\frac{2q}{m}}\frac{V^{\nicefrac{3}{2}}}{D^2}.
    \label{eq:J_CL}
\end{equation}
From this we may calculate the expected steady state current in our system 
\begin{equation}
    I_\mathrm{2D} =\frac{\pi}{9}\varepsilon_0\sqrt{\frac{2q}{m}}V^{\nicefrac{3}{2}}\left(\frac{4R^2}{D^2}+\frac{R}{D} \right).
    \label{eq:I_2DCL}
\end{equation}
 
For short-pulse emission a one-dimensional model (assuming a planar diode of infinite extent) predicts that for a pulse of constant injected current of duration $\tau_\mathrm{pulse}$ there is a critical current density, $J_\mathrm{crit}$ that is the maximum allowed to ensure that a virtual cathode does not form.
In its simplest form, where the bunch of charge is approximated as a single sheet, this critical current density is given by~\cite{valfells_effects_2002}
\begin{equation}
    J_\mathrm{crit} =\frac{\varepsilon_0 V}{\tau_\mathrm{pulse} D}.
    \label{eq:J_crit}
\end{equation}

This model essentially assumes that for a short bunch, approximated as a single sheet, a virtual cathode will form when the surface-charge density of the sheet is equal to the surface-charge density of the cathode surface due to the applied electric field, namely $\sigma = \varepsilon_0 V / D$.
The approximation that, under space-charge limited conditions, the charge may considered to be a single sheet, of the aforementioned surface charge density, transiting the gap in the appropriate time, is called the capacitive model and has been used successfully to derive the classic Child-Langmuir law~\cite{umstattd_simple_2005}.
The capacitive model for charge density has also been used for analysis of short beam bunches of finite diameter~\cite{bazarov_maximum_2009}.

\begin{figure}[!t]
    \centering
    \includegraphics[width=\linewidth]{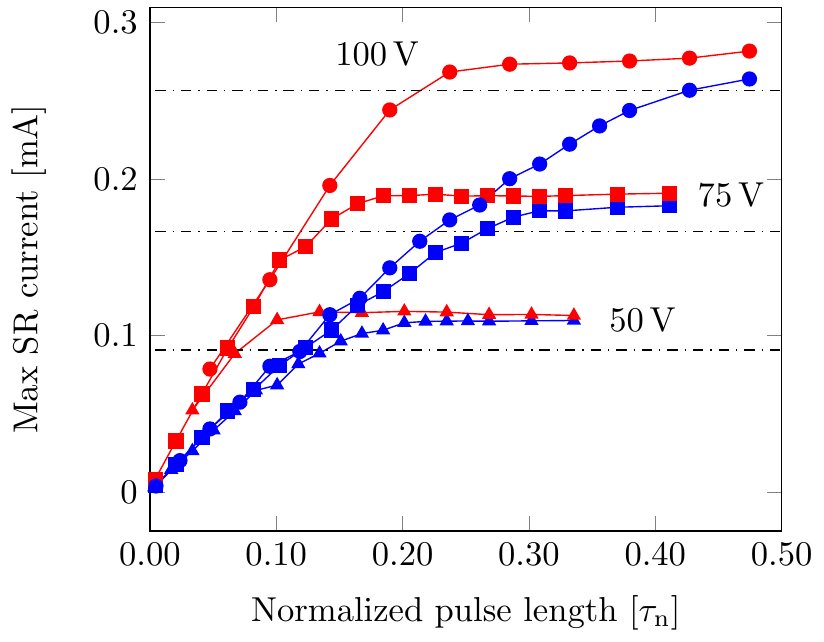}
    \caption{Induced current vs.\ normalized pulse length for three different gap voltages: 50, 75, 100 V, 250 nm emitter radius with curves for scaled laser amplitudes of 5 (blue) and 10 (red). The current from \autoref{eq:I_2DCL} is represented by the horizontal dash-dotted lines for 50, 75, 100 V.}
    \label{fig:I_tp_CL}
\end{figure}

\begin{figure}[!t]
    \centering
    \includegraphics[width=\linewidth]{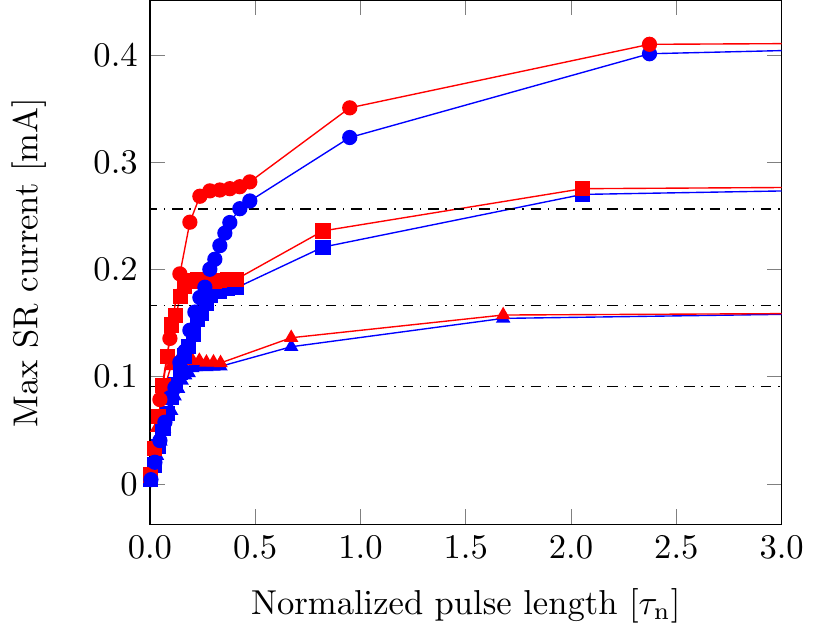}
    \caption{50, 75, 100 V, 250 nm emitter for scaled laser amplitude of 5 (blue) and 10 (red).: the CL limit from \autoref{eq:I_2DCL} is drawn in black for the different gap voltages.}
    \label{fig:long_I_tp_CL}
\end{figure}

Let us now look at the current as a function of the width of the laser pulse.
\autoref{fig:I_tp_CL} shows how the maximum value of the induced current varies with the normalized pulse length (the ratio of the laser pulse width to the transit time of a single electron across the diode gap).
We note that for very short pulse width the maximum current increases proportionally with the normalized pulse length.
This is indicative of source limited emission where charge can continually be added to the diode gap in proportion to the rate of photoemission and pulse duration.
The growth rate is independent of applied voltage but dependant on the amplitude of the laser pulse.
For slightly longer pulses, that are nonetheless short compared to the transit time, the accumulation of space-charge is sufficient to block further emission of photo-electrons and the current reaches a plateau indicative of space-charge limited emission.
This current limit is determined by the applied voltage, but note that for lower laser amplitude a greater normalized pulse length is needed to reach the critical amount of charge in the electron bunch.
We can also see that the space-charge limited current is somewhat higher than that predicted by \autoref{eq:I_2DCL}.
The space-charge limited current observed here corresponds to the maximum current shown within the circle in \autoref{fig:multi_ramo_current} for laser pulse width ranging from 
120 fs to 400 fs.
In \autoref{fig:long_I_tp_CL} the abscissa has been extended to show values of the normalized pulse width that extend beyond unity.
From this figure we can see the transition from the plateau that corresponds to space-charge limited current for a short bunch to another, higher, plateau that corresponds to the space-charge limited steady-state current.
Here the steady-state, space-charge limited current is considerably higher than predicted by \autoref{eq:I_2DCL}.
It has previously been observed~\cite{gunnarsson_space-charge_2021} that the steady-state space-charge limited current from microscopic emitters can deviate considerably from what is predicted by the simple 2D Child-Langmuir law, due to the relatively high contribution to the current from the edge of the emitting area, the so-called wing-structure of the emission profile~\cite{umstattd_two-dimensional_2001,luginsland_beyond_2002}, transverse beam expansion, and discrete particle effects.
Hence, the underestimation of \autoref{eq:I_2DCL} for the steady-state, space-charge limited current is not unexpected. 
What may seem unexpected, in light of previous work on short-pulse emission, is that the space-charge limited current for a short pulse is less than the space charge-limited current for the steady-state.
This is not a discrepancy.
In part the explanation lies with the fact that in the previous work on the short-pulse space-charge limit we are looking at the maximum injection current allowed so as not to form a virtual cathode over the duration of a given pulse length, whereas in our model we are looking at the induced current (which is due to the transit of the critical bunch of charge, once formed, across the diode gap).
Thus the inverse scaling of the critical current with pulse length is not appropriate for our purpose.
However, the reason that the space-charge limited current in the steady-state is higher than that of a short bunch has a physical reason associated with the mechanics of space-charge limited current from a microscopic emitter, as will be described when the total charge of the pulse is examined in the following paragraphs.

\begin{figure}[!t]
\centering
\includegraphics[width=\linewidth]{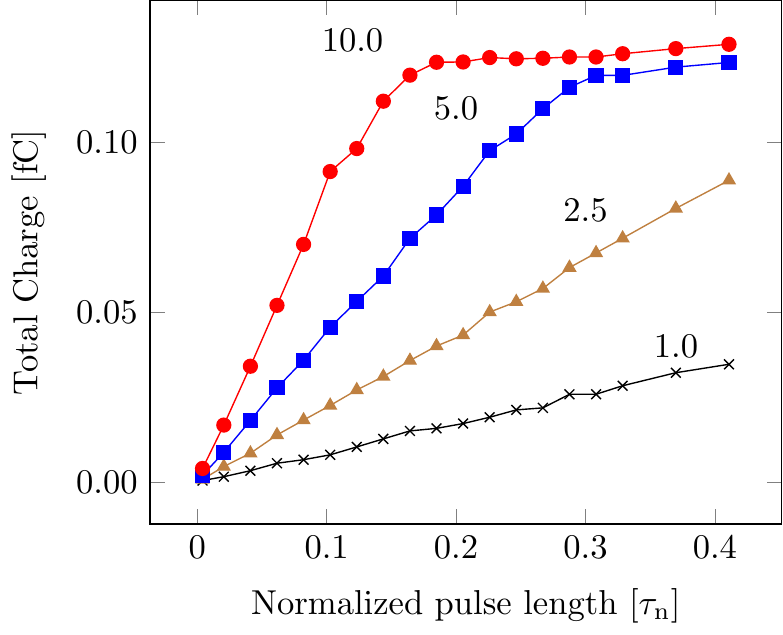}
    \caption{Total charge vs. normalized pulse length for scaled laser amplitude of 
1, 2.5, 5, 10. Gap voltage of 75 V, 250 nm radius emitter. The lowest amplitude (black) will not reach the space-charge limit before saturating the gap with constant current.}
    \label{fig:amplitude}
\end{figure}

Since it is apparent that, for short pulses, it is the single sheet model that is most appropriate, we now turn our attention to the amount of charge in the electron bunch as a function of laser pulse width for different values of scaled laser amplitude, emitter area and gap voltage.
\autoref{fig:amplitude} shows how the laser amplitude affects total charge as a function of the normalized pulse length.
Greater amplitude corresponds to a higher rate of photo-electrons being produced at the cathode.
The space-charge limit shows up as a plateau in the total charge in the pulse.
For low laser amplitudes the space-charge limit can not be reached, whereas it is obtained at shorter pulse lengths for larger amplitudes.

\begin{figure}[!t]
\centering
\includegraphics[width=\linewidth]{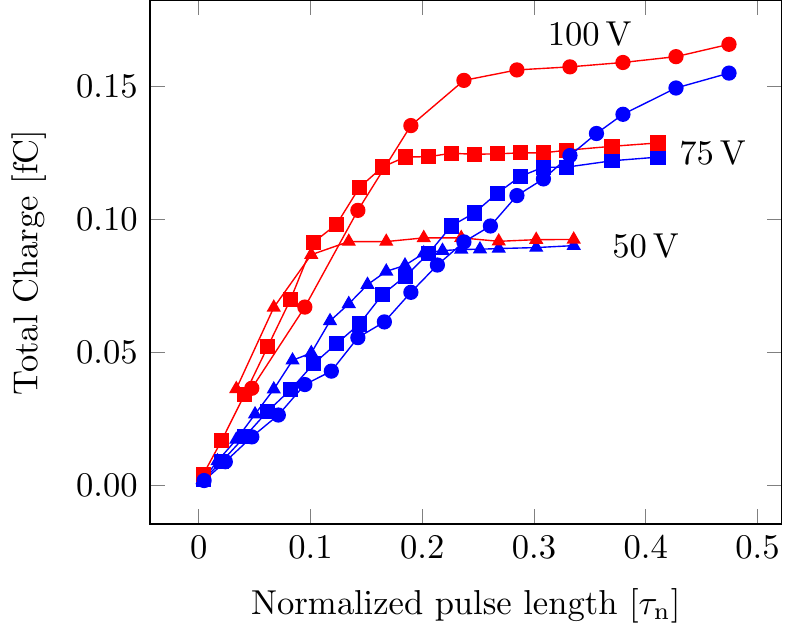}
    \caption{Total charge vs. normalized pulse length for three different voltages. 
50, 75, 100 V, 250 nm emitter, scaled laser amplitude of 5 (blue) and 10 (red). Increased voltage affects the space-charge limit with respect to pulse length while slope is related to amplitude.}
    \label{fig:voltage}
\end{figure}

From \autoref{fig:voltage} we can see how the gap voltage affects the total charge.
From the simple, single sheet model where the charge density of the sheet is given by $\sigma =\varepsilon_0 V / D$, and the emitter radius is 250 nm we would expect a total charge of 0.35 fC (femto Coulomb), 0.52 fC, and 0.70 fC for gap voltages of 
50 V, 75 V, and 100 V, respectively.
The measured charge is \numrange{2}{3} times higher.
The reason for this difference is that the estimation for the charge density of the sheet does not take into account the effects of limited emitter area that are implicit in the 2D Child-Langmuir law.
Note for instance that the space-charge limited current density from \autoref{eq:I_2DCL} with values of $R =250$ nm and $D =\ 2500$ nm is 3.5 times higher than the space-charge limited current density for an emitter of infinite extent.
From the single sheet model we would anticipate a linear relationship between the critical charge and the gap voltage, but, in fact the charge increases at a lower rate with voltage, e.g.\ the total charge for the pulse at 100 V is only 60 percent greater than the total charge at 50 Volt.
We do not have an explanation for this. 

\begin{figure}[!t]
\centering
\includegraphics[width=\linewidth]{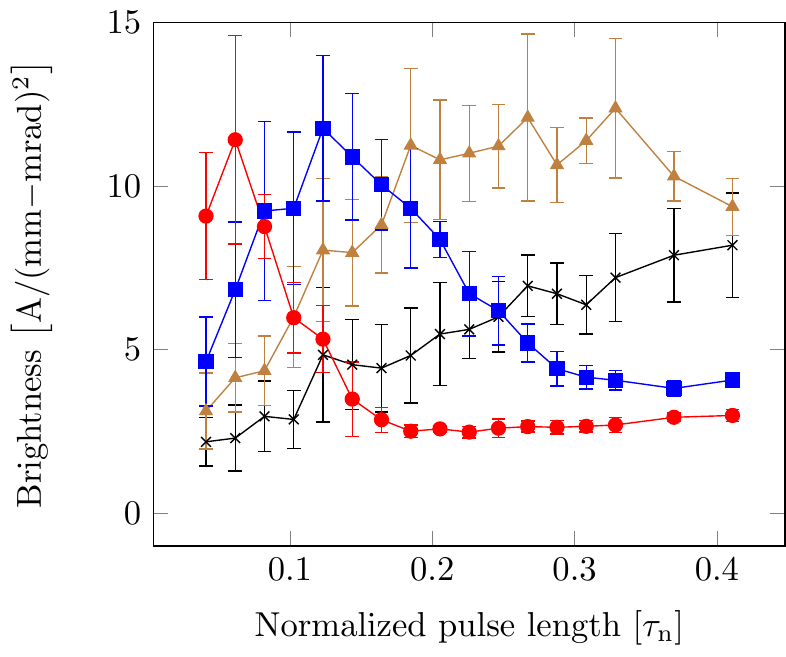}
\caption{Brightness versus normalized pulse length for different scaled laser amplitudes. 75 V gap potential, 250 nm emitter radius: The brightness peak shifts to the left, shorter pulse width, as the amplitude increases. Black, brown, blue, red, correspond to 
1, 2.5, 5, 10 in amplitude respectively.}
\label{fig:Br_tp_V}
\end{figure}

\begin{figure*}[!t]
\centering
\includegraphics[width=\linewidth]{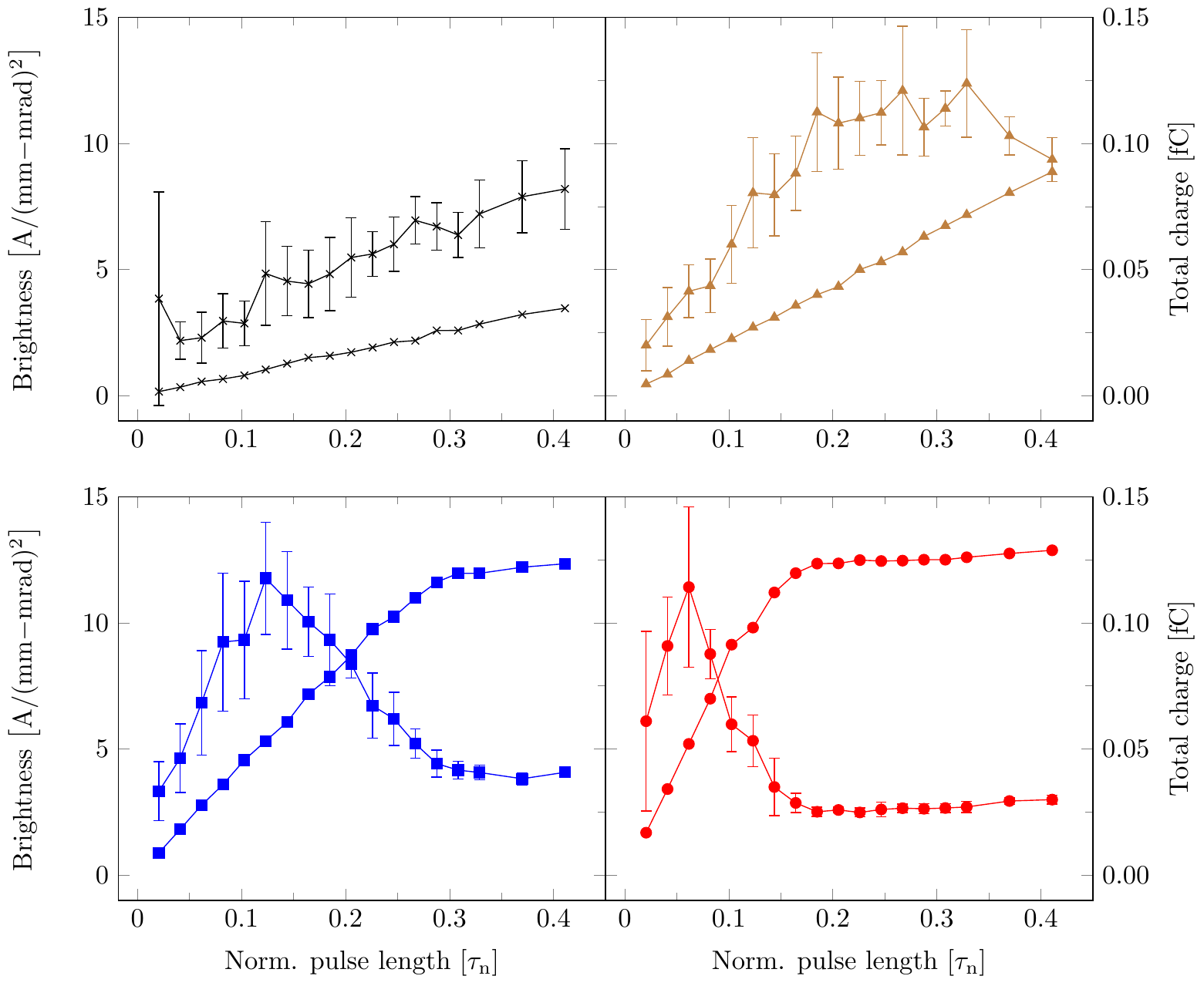}
\caption{Brightness and total pulse charge versus pulse length. 75 V gap potential, 
250 nm emitter radius: Charge and brightness lines of rising amplitudes have the same colors, brightness has error bars. Black, brown, blue, red, correspond to 
1, 2.5, 5, 10 in amplitude respectively. The Brightness peak hits at approx. 40\% of total charge peak.}
\label{fig:Br_Q_four}
\end{figure*}

Next we look at how the total charge is affected by the radius of the emitting area.
We see a linear rise in the pulse charge with pulse length until a plateau due to space-charge limitation is reached.
For the single sheet, capacitive, model we anticipate that the charge at the plateau scales with the area of the emitter (the emitter radius to the second power).
This is not the case as the current from the 250 nm radius emitter is roughly three times as high as that from the 125 nm radius emitter, rather than four times higher as might be expected for a single sheet of uniform charge density.
This can probably be explained by the fact that edge emission has a larger contribution to the total charge for the emitter of smaller radius.
We also note that for larger values of the normalized pulse, the total charge increases again as the transition from the single-sheet regime to the steady-state filled cathode regime begins.
Recall that the steady-state space-charge limited current is greater than that anticipated by \autoref{eq:I_2DCL} due to a large fraction of the emission coming from the edge of the emitter and due to transverse expansion of the beam~\cite{gunnarsson_space-charge_2021}.
This effect becomes more prominent as the ratio $Z_c / R$ increases, where $Z_c$ denotes the elevation of the center of charge above the cathode and $R$ is the emitter radius.
As a result of this, the steady-state current transition to the space-charge regime for larger area emitters begins when the beam bunch has propagated further along the diode than for smaller emitters.
Hence, the transition between regimes occurs earlier for smaller emitters.
\begin{figure}[!t]
\centering
\includegraphics[width=\linewidth]{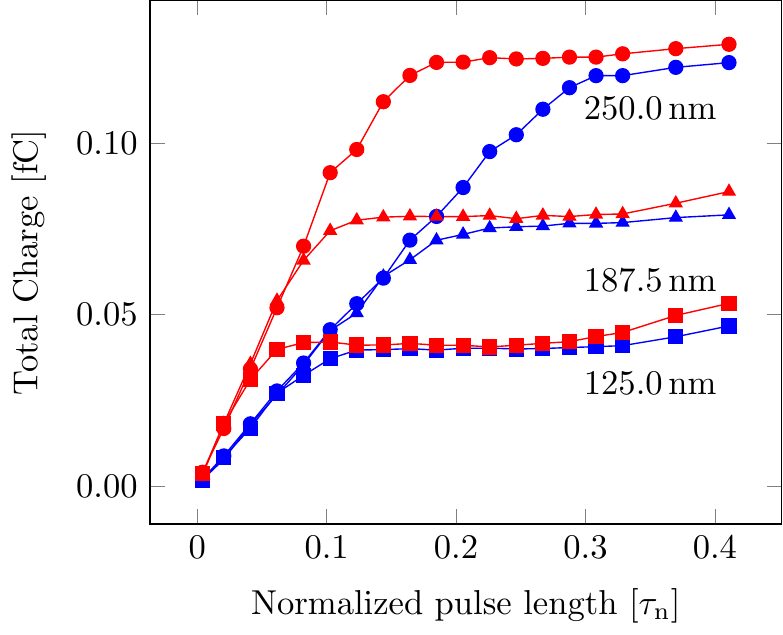}
    \caption{75 V, 125, 187.5, 250 nm radius emitters, 5, 10 Amplitude. Emitter size increases the space-charge limit with respect to pulse length while again the slope is related to amplitude.}
    \label{fig:emitter_size}
\end{figure}
Finally, we turn our attention to the brightness of the electron beam.
We look at the data underlying \autoref{fig:amplitude} and plot the brightness of the beam as a function of the normalized pulse length for different values of the laser amplitude.
This can be seen in \autoref{fig:Br_tp_V} and \autoref{fig:Br_Q_four} where a peak value of the brightness is apparent.
The peak value is roughly constant, though the peak becomes sharper as the scaled laser amplitude increases.
We note that the peak brightness is achieved when the charge in the electron bunch corresponds to roughly 40 percent of the critical charge for the short-pulse space-charge limit, which holds true for all voltages and emitter radii that were simulated.

\section{Summary and conclusions}
Using MD-Simulations we examined the transition from source limited emission to space-charge limited emission in photo-emitted electron beams in a microscopic diode for different values of laser pulse width, intensity, emitter area (or spot size) and accelerating potential.
We found that conventional capacitive models of short-pulse electron bunches may considerably underestimate their total charge due to neglecting two-dimensional space-charge effects, whereas the estimates for the short-pulse space-charge limited current using the approach of Koh and Ang, shown in \autoref{eq:Koh}, give a reasonably accurate estimate for the maximum current induced by a electron bunch transiting the diode gap.
We also identified parameters for optimal brightness of the beam bunch.
For the parameter range that we studied it is found that the highest value of brightness occurs when the charge in the beamlet is roughly 40 percent of the critical charge for formation of a virtual cathode.
This result is similar to what has been found for thermal emission in microdiodes, in that optimal brightness is achieved at a point during transition from source-limited to space-charge limited emission.
This may have some practical value for designers of electron sources for coherent and time resolved electron beams.

\section*{Acknowledgment}
This material is based upon work supported by the Icelandic Research Fund grant number 174512-051 and the Reykjavik University doctoral fund.

\bibliographystyle{IEEEtran}

\end{document}